\documentclass[a4paper,nobibnotes,nofootinbib]{revtex4}
\usepackage{amssymb}
\usepackage{amsmath}
\usepackage{epsfig}
\newcommand{\be}{\begin{equation}}
\newcommand{\ee}{\end{equation}}
\newcommand{\bea}{\begin{eqnarray}}
\newcommand{\eea}{\end{eqnarray}}
\newcommand{\nn}{\nonumber}

\newcommand{\bs}{\boldsymbol}
\newcommand{\g}{\gamma}
\newcommand{\f}{\frac}

\newcommand{\xp}{x_{\mathbb P}}

\newcommand\lr[1]{{\left({#1}\right)}}

\newcommand{\PO}{\rm l \! P }

\newcommand{\xpom}{x_{\PO} }

\begin{document}

\title{Geometric scaling in diffractive deep inelastic scattering}

\author{C. Marquet}\email{marquet@spht.saclay.cea.fr}
\affiliation{Service de Physique Th{\'e}orique, CEA/Saclay,
91191 Gif-sur-Yvette cedex, France\\
URA 2306, unit{\'e} de recherche associ{\'e}e au CNRS}

\author{L. Schoeffel}\email{schoffel@hep.saclay.cea.fr}
\affiliation{DAPNIA/Service de physique des particules, CEA/Saclay, 91191 
Gif-sur-Yvette cedex, France}

\begin{abstract}

The geometric scaling property observed in the HERA data at small $x,$ that the 
deep inelastic scattering (DIS) total cross-section is a function of only the 
variable $Q^2 x^\lambda,$ has been argued to be a manifestation of the 
saturation regime of QCD and of the saturation scale $Q^2_s(x)\!\sim\! 
x^{-\lambda}.$ We show that this implies a similar scaling in the context of 
diffractive DIS and we observe, for several diffractive observables, that the 
experimental data from HERA confirm the expectations of this scaling.

\end{abstract}
\maketitle

\section{Introduction}

Deep inelastic scattering (DIS) is a process in which a virtual photon is used 
as a hard probe to resolve the small distances inside a proton and study its 
partonic constituents: quarks and gluons that obey the laws of perturbative 
QCD. When probing with a fixed photon virtuality $Q^2\!\gg\!\Lambda_{QCD}^2,$ 
and increasing the energy of the photon-proton collision $W$, the parton 
densities inside the proton grow. Eventually, at some energy much bigger than 
the hard scale, corresponding to a small value of $x\!=\!Q^2/W^2$, one enters 
the saturation regime of QCD~\cite{glr}: the gluon density becomes so large that 
non-linear effects like gluon recombination become important, taming the growth 
of the parton densities.

The transition to the saturation regime is characterized by the so-called 
saturation momentum $Q_s(x)\!=\!Q_0\ x^{-\lambda/2}.$ This is an intrinsic scale 
of the high-energy proton which increases as $x$ decreases. 
$Q_0\!\sim\!\Lambda_{QCD},$ but as the energy increases, $Q_s$ becomes a hard 
scale, and the transition to saturation occurs when $Q_s$ becomes comparable to 
$Q.$ The higher $Q^2$ is, the smaller $x$ should be to enter the saturation 
regime. 

Although the saturation regime is only reached when $Q_s\!\sim\!Q,$ 
observables are sensitive to the saturation scale already during the 
approach to saturation~\cite{extscal} when $\Lambda_{QCD}\!\ll\!Q_s\!\ll\!Q.$ 
For inclusive events in deep inelastic scattering, this feature manifests itself 
via the so-called geometric scaling property: instead of being a function of 
$Q^2/Q_0^2$ and $x$ separately, the total cross-section is only a function of 
$\tau\!=\!Q^2/Q_s^2(x),$ up to large values of $\tau.$ Experimental measurements 
of inclusive DIS are compatible with that prediction~\cite{geoscal}.

Part of the DIS events are diffractive, meaning that the proton remains intact 
after the collision and that there is a rapidity gap between that proton and 
the rest of the final-state particles. Such events are expected to be much more 
sensitive to the saturation regime of QCD than the inclusive ones~\cite{golec} 
and our goal in this letter is to extend the geometric scaling property to 
diffractive DIS and to compare the resulting predictions with the available 
experimental data. We shall consider several diffractive observables, focusing 
on inclusive hard diffraction and vector-meson production.

In the saturation regime of QCD, contributions to the cross-sections growing 
like $Q_s/Q$ are important. The leading-twist approximation of perturbative QCD, 
in which $Q^2$ is taken as the bigger scale, cannot account for such 
contributions, and therefore is not appropriate to describe the small-x limit of 
deep inelastic scattering. As leading-twist gluon distributions cannot be used 
to compute cross-sections, the dipole picture of DIS~\cite{dipole} has been 
developed to describe the high-energy limit. It expresses the hadronic 
scattering of the virtual photon through its fluctuation into a color singlet 
$q\bar q$ pair (or dipole) of a transverse size $r\!\sim\!1/Q$. The dipole is 
then the hard probe that resolves the small distances inside the proton. 

The dipole picture naturally incorporates the description of both inclusive and 
diffractive events into a common theoretical framework~\cite{nikzak,biapesroy}, 
as the same dipole scattering amplitudes enter in the formulation of the 
inclusive and diffractive cross-sections. This will be recalled in Section II 
and will allow us to extend the geometric scaling property to diffractive 
observables, as detailed in Section III. Finally, Section IV is devoted to 
comparison with experimental data and Section V concludes.

\section{The dipole picture of deep inelastic scattering}

We focus on diffractive DIS: $\g^*p\!\rightarrow\!Xp.$ The proton gets out of 
the $\g^*\!-\!p$ collision intact, and there is a rapidity gap between that 
proton and the final state $X$ whose invariant mass we denote $M_X.$ We recall 
that the photon virtuality is denoted $Q^2,$ and the $\g^*\!-\!p$ total energy 
$W.$ It is convenient to introduce the following variables:
\be
x=\f{Q^2}{Q^2+W^2}\ ,\hspace{1cm}\beta=\f{Q^2}{Q^2+M_X^2}\ 
,\hspace{1cm}\xp=x/\beta\ 
.\ee
The $\g^*\!-\!p$ total cross-section $\sigma^{\g^*p\rightarrow X}_{tot}$ is 
usually expressed as a function of $x$ and $Q^2,$ while the diffractive 
cross-section $d\sigma^{\g^*p\rightarrow Xp}_{diff}/d\beta$ is expressed as a 
function of $\beta,$ $\xp,$ and $Q^2.$ Note that the size of the rapidity gap in 
the final state is $\ln(1/\xp).$

To compute those cross-sections in the high-energy limit, it is convenient to 
view the process in a particular frame called the dipole frame. In this frame, 
the virtual photon undergoes the hadronic interaction via a fluctuation into a 
colorless $q\bar q$ pair, called dipole, which then interacts with the target 
proton. The wavefunctions $\psi_\lambda^{f,\alpha\beta}(z,\textbf{r};Q^2)$ 
describing the splitting of a virtual photon with polarization $\lambda$ into a 
dipole are well known. The indices $\alpha$ and $\beta$ denote the spins of the 
quark and the antiquark composing the dipole of flavor $f.$ The wavefunctions 
depend on $Q^2,$ the fraction $z$ of longitudinal momentum (with respect to the 
$\g^*\!-\!p$ collision axis) carried by the quark, and the two-dimensional 
vector $\textbf{r}$ whose modulus is the transverse size of the dipole 
(transverse coordinates are obtained from a Fourier transform of transverse 
momenta). Formulae giving the funcions $\psi_\lambda^{f,\alpha\beta}$ can be 
found in the literature (see for instance~\cite{biapesroy}). In what follows, we 
will need the functions $\Phi^f_\lambda$ which describe the overlap between two 
wavefunctions for splitting into dipoles of different transverse size 
$\textbf{r}$ and $\textbf{r}':$
\be
\phi^f_\lambda(z,\textbf{r},\textbf{r}';Q^2)=N_c\sum_{\alpha\beta}
\left[\psi_\lambda^{f,\alpha\beta}(z,\textbf{r}';Q^2)\right]^*
\psi_\lambda^{f,\alpha\beta}(z,\textbf{r};Q^2)\ .\label{overlap}
\ee
For a transversely (T) or longitudinally (L) polarized photon, these functions 
are given by
\be
\phi^f_T(z,\textbf{r},\textbf{r}';Q^2)=
\frac{\alpha_{em}N_c}{2\pi^2}e_f^2
\left((z^2+(1-z)^2)\varepsilon_f^2
\f{\textbf{r}.\textbf{r}'}{|\textbf{r}||\textbf{r}'|}
K_1(\varepsilon_f|\textbf{r}|)K_1(\varepsilon_f|\textbf{r}|')
+m_f^2 K_0(\varepsilon_f|\textbf{r}|)
K_0(\varepsilon_f|\textbf{r}|')\right)\ ,\ee
\be
\phi^f_L(z,\textbf{r},\textbf{r}';Q^2)=
\frac{\alpha_{em}N_c}{2\pi^2}e_f^2
4Q^2 z^2(1-z)^2 K_0(\varepsilon_f|\textbf{r}|)
K_0(\varepsilon_f|\textbf{r}|')\ .\ee
In the above, $e_f$ and $m_f$ denote the charge and mass of the quark with 
flavor $f$ and $\varepsilon_f^2\!=\!z(1\!-\!z)Q^2\!+\!m_f^2.$ 

\textbf{The total cross-section} $\sigma^{\g^*p\rightarrow 
X}_{tot}.$ Via the optical theorem, it is related to the elastic scattering of 
the virtual photon. In the dipole frame, this scattering happens as follows: at 
a given impact parameter $\textbf{b},$ the photon splits into a dipole with a 
given size $\textbf{r}$ which scatters elastically off the proton and recombines 
back into the photon. Therefore the overlap function $\Phi_\lambda^{\g^*\g^*}$ 
which enters in the computation of the total cross-section is
\be
\Phi_\lambda^{\g^*\g^*}(z,|\textbf{r}|;Q^2)=\sum_f 
\phi^f_\lambda(z,\textbf{r},\textbf{r};Q^2)\ .
\ee
The total cross-section is then given by (for fixed impact parameter 
$\textbf{b}$):
\be
\f{d\sigma^{\g^*p\rightarrow X}_{tot}}{d^2b}(x,Q^2)=2
\int d^2r \int_0^1 dz
\sum_{\lambda=L,T}\Phi_\lambda^{\g^*\g^*}(z,|\textbf{r}|;Q^2)\
T_{q\bar q}(\textbf{r},\textbf{b};x)\label{tot}\ee
where the function $T_{q\bar q}(\textbf{r},\textbf{b};x)$ is the elastic 
scattering amplitude of the dipole of size $\textbf{r}$ off the proton at impact 
parameter $\textbf{b}.$ It contains the $x$ dependence, reflecting the fact that 
in our frame, the proton carries all the energy and is therefore evolved up to 
the rapidity $\ln(1/x).$ In the high-energy limit $x\!\ll\!1$ we are considering 
here, $T_{q\bar q}$ does not depend on $z.$

\textbf{The diffractive cross-section} $d\sigma^{\g^*p\rightarrow 
Xp}_{diff}/d\beta.$ In the amplitude, the photon splits into a dipole of size 
$\textbf{r}$ which scatters off the proton and dissociates into a final state
of invariant mass $M_X.$ The same happens in the complex conjugate amplitude, 
except that the dipole size $\textbf{r}'$ is different from $\textbf{r}.$
Indeed, the final state is characterized by a particular value of $M_X$ (or 
equivalently $\beta$), corresponding to a particular momentum of the 
quark-antiquark pair. In coordinate space, this imposes two different dipole 
sizes in the amplitude and the complex conjugate amplitude, therefore the 
functions $\phi_\lambda^f(z,\textbf{r},\textbf{r}';Q^2)$ (see \eqref{overlap}) 
enter in the computation of the diffractive cross-section:
\bea
\f{d\sigma^{\g^*p\rightarrow Xp}_{diff}}{d^2b\ d\beta}(\beta,\xp,Q^2)=
\frac{Q^2}{4\pi\beta^2}\sum_f \int d^2r \int d^2r'
\int_0^1 dz z(1-z) \Theta(\bs{\kappa}_f^2)\
e^{i\bs{\kappa}_f.(\textbf{r}'-\textbf{r})}\hspace{3cm}\nn\\
\sum_{\lambda=L,T}\phi_\lambda^f(z,\textbf{r},\textbf{r}';Q^2)
T_{q\bar q}(\textbf{r},\textbf{b};\xp)
T_{q\bar q}(\textbf{r}',\textbf{b};\xp)\ .\label{diff}\eea
In the above, $\bs{\kappa}_f^2\!=\!z(1\!-\!z)Q^2(1\!-\!\beta)/\beta\!-\!m_f^2.$ 
Note that now, the proton is only evolved up to the rapidity $\ln(1/\xp).$ This 
is because some of the energy ($M_X^2$) is carried by the dipole in order to 
form the diffractive final state. The dipole is evolved up to a rapidity 
$\ln(1/\beta)$ and the proton up to the rapidity $\ln(\beta/x)\!=\!\ln(1/\xp).$ 
The high-energy limit in this case is $\xp\!\ll\!1.$

To write formula \eqref{diff}, we have neglected possible final 
states containing gluons. This is justified because these are suppressed by 
extra powers of $\alpha_s.$ However, if $\beta$ becomes too small, the dipole 
evolves to higher rapidities and emits soft gluons. Large logarithms 
$\alpha_s\ln(1/\beta)$ coming from the emission of those soft gluons arise, and 
multiple gluons emissions should be resummed to complete formula \eqref{diff}.
These multiple gluon emissions in the limit $\beta\ll1$ can also be accounted 
for in the dipole picture~\cite{diffscal}, provided one uses the large$-N_c$ 
limit. Indeed in this limit, a dipole emitting a soft gluon is equivalent to a 
dipole splitting into two dipoles. To illustrate this, the contribution of the 
$q\bar qg$ final state reads (see also~\cite{gbar,gkop,gkov,gmun,gmar}):
\be
\f{d\sigma^{\g^*p\rightarrow (X\!=\!q\bar qg)p}_{diff}}{d^2b\ d\beta}
=\frac{\alpha_s N_c}{2\pi^2\beta}\int d^2r \int_0^1 dz
\sum_{\lambda=L,T}\Phi_\lambda^{\g^*\g^*}(z,|\textbf{r}|;Q^2)\int d^2z 
\f{\textbf{r}^2}{\textbf{z}^2(\textbf{r}\!-\!\textbf{z})^2}
\Big[T^{(2)}_{q\bar q}(\textbf{z},\textbf{r}\!-\!\textbf{z},\textbf{b};\xp)
-T_{q\bar q}(\textbf{r},\textbf{b};\xp)\Big]^2\ .\label{diffqqg}\ee
In the above, the function $T^{(2)}_{q\bar 
q}(\textbf{z},\textbf{r}\!-\!\textbf{z},\textbf{b};\xp)$ is the scattering 
amplitude for two dipoles of sizes $\textbf{z}$ and 
$\textbf{r}\!-\!\textbf{z}$ at impact parameters 
$\textbf{b}\!-\!(\textbf{r}\!-\!\textbf{z})/2$ and $\textbf{b}\!-\!\textbf{z}/2$ 
respectively. These come from the splitting of the dipole of size $\textbf{r}$ 
at impact parameter $\textbf{b}.$ Moreover the overlap function is 
$\Phi_\lambda^{\g^*\g^*}.$ It is so because in the leading $\ln(1/\beta)$ 
approximation, the final state mass $M_X$ is fixed only by the soft gluon 
longitudinal momentum, and therefore transverse sizes are the same in the 
amplitude and the complex conjugate amplitude\footnote{Let us also mention 
another approach to compute the $q\bar qg$ contribution in the dipole 
picture, which resums logarithms of $Q^2$ instead of logarithms of $1/\beta.$ In 
this case, it is an effective gluonic dipole which scatters off the 
proton~\cite{gwus}.}.

\textbf{The diffractive vector-meson production cross-section}
$\sigma^{\g^*p\rightarrow Vp}_{VM}.$ In this case, after scattering off the 
proton, the dipole recombines into a particular final state $X\!=\!V$, a vector 
meson whose mass we shall denote $M_V.$ To describe this process, we need to 
introduce the wavefunction $\varphi_\lambda^{f,\alpha\beta}(z,\textbf{r};M_V^2)$ 
which describes the splitting of the vector meson with polarization $\lambda$ 
into the dipole. The overlap function $\Phi^{\g^*V}_{\lambda}$ which enters in 
the computation of the vector-meson production amplitude is then given by
\be
\Phi^{\g^*V}_\lambda(z,|\textbf{r}|;Q^2,M_V^2)=\sum_{f\alpha\beta}
\left[\varphi_\lambda^{f,\alpha\beta}(z,\textbf{r};M_V^2)\right]^*
\psi_\lambda^{f,\alpha\beta}(z,\textbf{r};Q^2)\ .
\ee
Because the final state has been explicitely projected into a vector meson, the 
two dipoles in the amplitude and the complex conjugate amplitude are not 
connected and the cross-section (for fixed impact parameter) is simply the 
square of the amplitude:
\be
\f{d\sigma^{\g^*p\rightarrow Vp}_{VM}}{d^2b}(\xp,Q^2)=\left|
\int d^2r \int_0^1 dz
\sum_{\lambda=L,T}\Phi_\lambda^{\g^*V}(z,|\textbf{r}|;Q^2,M_V^2)\
T_{q\bar q}(\textbf{r},\textbf{b};\xp)\right|^2\ .\label{vm}
\ee
The functions $\Phi^{\g^*V}_{\lambda}$ depend on the meson wavefunctions 
$\varphi_\lambda^{f,\alpha\beta},$ and several different models exist in the 
literature~\cite{mwfs1,mwfs2,mwfs3}. As emphasized later, we shall only use 
model-independent features of $\Phi^{\g^*V}_{\lambda}.$ Formula \eqref{vm} can 
be used to compute the Deeply-Virtual-Compton-Scatering (DVCS) cross-section 
$\sigma^{\g^*p\rightarrow \g p}_{DVCS},$ provided one uses the overlap function 
\be
\Phi^{\g^*\g}_T(z,|\textbf{r}|;Q^2)=\sum_{f\alpha\beta}
\left[\psi_T^{f,\alpha\beta}(z,\textbf{r};0)\right]^*
\psi_T^{f,\alpha\beta}(z,\textbf{r};Q^2)
\ee
between the incoming virtual photon and the outgoing transversely-polarized real 
photon.

\section{Saturation, geometric scaling and its consequences}

Using the dipole picture of deep inelastic scattering, we have expressed the 
total \eqref{tot}, diffractive \eqref{diff} and exclusive vector-meson 
production \eqref{vm} cross-sections in the high-energy limit in terms of a 
single object: the dipole scattering amplitude off the proton $T_{q\bar 
q}(\textbf{r},\textbf{b};x).$ It is mainly a non-perturbative quantity, but its 
evolution towards small values of $x$ (or high energy) is computable from 
perturbative QCD. Evolution equations have been established in the leading 
$\ln(1/x)$ approximation~\cite{bk,jimwlk} and, at least for central impact 
parameter, one has learned a lot about the growth of the dipole amplitude and 
the transition from the leading-twist regime $T_{q\bar q}\!\ll\!1$ towards and 
into the saturation regime $T_{q\bar q}\!=\!1.$

The most important prediction was probably the geometric scaling 
regime~\cite{extscal,tw}: at small values of $x,$ instead of being a function of 
a priori the two variables $\textbf{r}$ and $x,$ 
$T_{q\bar q}(\textbf{r},\textbf{b}\!\simeq\!0;x)$ is actually a function of the 
single variable $\textbf{r}^2Q^2_s(x)$ up to inverse dipole sizes significantly 
larger than the saturation scale $Q_s(x).$ In formulae, one can write
\be
T_{q\bar q}(\textbf{r},\textbf{b};x)=S(\textbf{b})\ T(\textbf{r}^2Q^2_s(x))
\label{scal}\ee
where we have introduced the impact parameter profile $S(\textbf{b}).$ 
Typically, $S(\textbf{b})\!=\!e^{-\textbf{b}^2/R_p^2}$ where $R_p$ is the 
transverse radius of the proton. When performing the $\textbf{b}$ integration in 
formulae \eqref{tot}, \eqref{diff} or \eqref{vm}, this contributes only to the 
normalization via a constant factor $\int d^2b\ S(\textbf{b})\!=\!S_p,$ 
caracterising the transverse area of the proton. If $\textbf{r}^2Q^2_s\!>\!1$ 
then $T\!=\!1$ and the scaling is obvious. We insist that the scaling property 
\eqref{scal} is a non-trivial prediction for $\textbf{r}^2Q^2_s\!\ll\!1,$ when 
$T$ is still much smaller that 1. Of course the geometric scaling window has a 
limited extension: at very small dipole sizes, deep into the leading-twist 
regime, the scaling breaks down. Universal scaling violations~\cite{tw} due to 
$x$ not being small enough have also been derived. Recently, a new type of 
scaling violations has been predicted~\cite{ploop,diffscal}, this one eventually 
arising when $x$ becomes even smaller, transforming the geometric scaling regime 
into an intermediate energy regime. 

In this work, we shall consider the case of exact scaling \eqref{scal}. As 
already mentioned, the resulting prediction for the DIS total cross-section is 
in very good agreement with experimental measurements~\cite{geoscal} (see 
also~\cite{schild}). Our goal is to further 
test the geometric scaling regime by considering its prediction for diffractive 
observables. But, as a reminder, let us start with the total cross-section. 
Neglecting quark masses, one can rewrite the cross-section \eqref{tot} as
\be
\sigma^{\g^*p\rightarrow X}_{tot}(x,Q^2)=2S_p\frac{\alpha_{em}N_c}{\pi}\sum_f 
e_f^2 \int_0^\infty \bar{r}d\bar{r}\int_0^1 dz
\left\{f_T(z)K_1^2(\sqrt{z(1\!-\!z)}\bar{r})
+f_L(z)K_0^2(\sqrt{z(1\!-\!z)}\bar{r})\right\}
T\lr{\f{Q^2_s(x)}{Q^2}\ \bar{r}^2}
\ee
where we have introcuced the functions $f_T(z)\!=\!(z^2+(1\!-\!z)^2)z(1\!-\!z)$ 
and $f_L(z)\!=\!4z^2(1\!-\!z)^2$ and rescaled the size variable $|\textbf{r}|$ 
to the dimensionless variable $\bar{r}\!=\!Q|\textbf{r}|.$ We obtain the 
geometric scaling of the total cross-section at small $x:$
\be
\sigma^{\g^*p\rightarrow X}_{tot}(x,Q^2)=
\sigma^{\g^*p\rightarrow X}_{tot}(\tau)\ ,\hspace{1cm}
\tau=Q^2/Q_s^2(x)\ .\label{test0}\ee
This has been seen confirmed by experimental data~\cite{geoscal} with $Q_s(x)$ 
given by
\be
Q_s(x)=Q_0\lr{\f{x}{x_0}}^{-\lambda/2}\ ,
\hspace{0.5cm}Q_0\equiv 1 \mbox{GeV}\label{qs}
\ee
and the parameters $\lambda\!=\!0.288$ and $x_0\!=\!3.04\ 10^{-4}$ taken 
from~\cite{golec}. In order to illustrate it, Fig.1 is an update of the original 
plot which shows the cross-section $\sigma^{\g^*p\rightarrow X}_{tot}$ as a 
function of $\tau$ with the latest data of the different experiments which 
provide measurements at $x\!<\!0.01:$ the H1~\cite{h1f2}, ZEUS~\cite{zeusf2}, 
E665~\cite{e665f2} and NMC~\cite{nmcf2} collaborations. Except for one E665 
point, the data do lie on the same curve. This is even true at low values of 
$Q^2,$ for which one could have expected scaling violations do to the charm 
quark mass, but one can see on Fig.1 that these are not sizable (see 
also~\cite{gonmac}).
\begin{figure}[ht]
\begin{center}
\epsfig{file=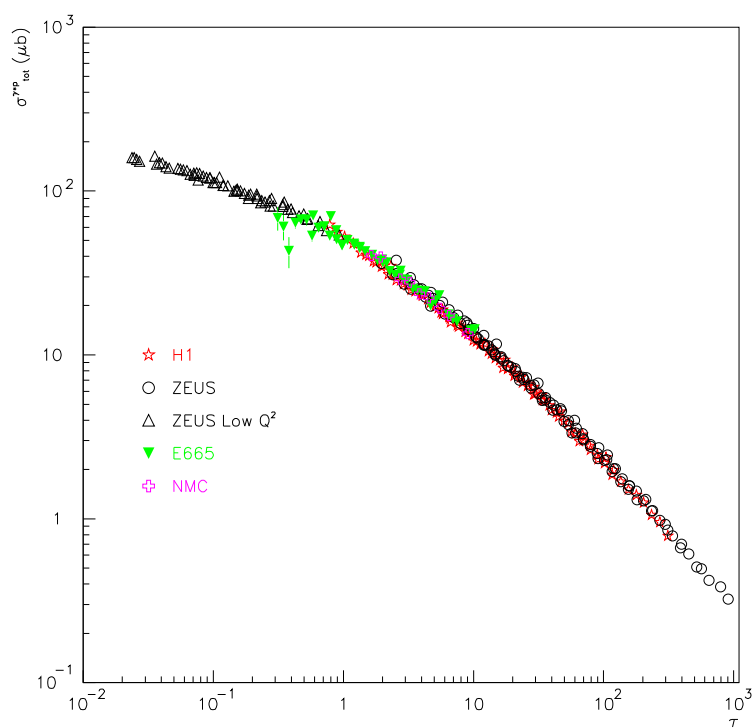,width=10cm}
\caption{The total cross section $\sigma^{\g^*p\rightarrow X}_{tot}$ as a 
function of $\tau$ for $x\!<\!0.01.$ The data are the most recent by the H1, 
ZEUS, E665 and NMC collaborations. Only statistical uncertainties are shown.}
\end{center}
\label{F1}
\end{figure}

Let us now consider the diffractive cross-section \eqref{diff}. It can be 
rewritten
\be
\f{d\sigma^{\g^*p\rightarrow Xp}_{diff}}{d\beta}(\beta,\xp,Q^2)=S_p
\f{\alpha_{em}N_c}{2\pi\beta^2}\sum_f e_f^2 \int_0^1 dz z(1-z)
\sum_{\lambda=L,T} f_\lambda(z)I^2_\lambda(z,\beta,Q^2_s(\xp)/Q^2)
\ee
with the following integral
\be
I_{T,L}(z,\beta,Q^2_s/Q^2)=\int_0^\infty \bar{r}d\bar{r} 
K_{1,0}(\sqrt{z(1\!-\!z)}\bar{r})
J_{1,0}(\sqrt{z(1\!-\!z)(1\!-\!\beta)/\beta}\bar{r})
T\lr{\f{Q^2_s}{Q^2}\ \bar{r}^2}
\ee
where $I_T$ contains $K_1$ and $J_1$ bessel functions and $I_L$ contains $K_0$ 
and $J_0.$ So, another signature of the saturation regime of QCD should then be 
the geometric scaling of the diffractive cross-section at fixed $\beta$ and 
small $\xp:$
\be
\f{d\sigma^{\g^*p\rightarrow Xp}_{diff}}{d\beta}(\beta,\xp,Q^2)=
\f{d\sigma^{\g^*p\rightarrow Xp}_{diff}}{d\beta}(\beta,\tau_d)\ 
,\hspace{1cm}\tau_d=Q^2/Q_s^2(\xp)\ .
\label{test1}
\ee
This will be discussed in the next Section. Note that for small values of 
$\beta,$ when using a more complete formulation of the diffractive 
cross-section, the prediction \eqref{test1} remains unchanged. For instance, in 
the geometric scaling regime, the behavior of $T^{(2)}_{q\bar q}$ is 
\be
T^{(2)}_{q\bar q}(\textbf{z},\textbf{r}\!-\!\textbf{z},\textbf{b};x)
=S(\textbf{b})
\tilde{T}(\textbf{z}^2Q^2_s(x),(\textbf{r}\!-\!\textbf{z})^2Q^2_s(x))\ee
and therefore when using formula \eqref{diffqqg}, the prediction \eqref{test1}
remains true and it also the case in other approaches~\cite{diffscal,gwus}.

Let us finally discuss vector-meson production. In this case, the problem is not 
as simple because of the extra scale $M_V$ and of the model-dependent meson 
wavefunctions. However, it is possible to take advantage of a quite general
feature rather independent of the particular model for the meson wavefunction:
longitudinally-polarized vector mesons are predominant and the overlap function 
follows the approximate scaling law (see for instance~\cite{mms})
\be
2\pi r^2\ \int_0^1 dz\ \Phi_L^{\g V}(z,\textbf{r};Q^2,M_V^2)\simeq 
g\lr{\textbf{r}^2(Q^2+M_V^2)}\label{mwfap}
\ee
where the function $g$ is sharply picked around 1. As a consequence of this, in 
the geometric scaling regime the vector-meson production cross-section can be 
rewritten
\be
\sigma^{\g^*p\rightarrow Vp}_{VM}(\xp,Q^2)=S_p\left|
\int_0^\infty \f{d\bar{r}}{\bar{r}} g(\bar{r}^2)
T\lr{\f{Q^2_s(\xp)}{Q^2+M_V^2}\bar{r}^2}\right|^2
\ee
and the following behavior can be predicted:
\be
\sigma^{\g^*p\rightarrow Vp}_{VM}(\xp,Q^2)=
\sigma^{\g^*p\rightarrow Vp}_{VM}(\tau_V)\ 
,\hspace{1cm}\tau_V=(Q^2+M_V^2)/Q_s^2(\xp)\ .\label{test2}
\ee
Note that, for the DVCS cross-section, the prediction is 
$\sigma^{\g^*p\rightarrow\g p}_{DVCS}(x,Q^2)\!=\!
\sigma^{\g^*p\rightarrow\g p}_{DVCS}(\tau)$ without relying on \eqref{mwfap}.

\section{Geometric scaling and diffractive observables}

We shall now test the predictions \eqref{test1} et \eqref{test2}. The H1 and 
ZEUS experiments at HERA have measured the diffractive cross section for the 
process $ep\!\rightarrow\!eXY$, selecting events with a large rapidity gap 
between the systems $X$ and $Y$ in case of H1~\cite{f2d97}, and using the 
so-called $M_X$-method in case of ZEUS~\cite{zeus}. $Y$ represents the scattered
proton, either intact of in a low-mass excited state, with $M_Y\!<\!1.6\ 
\mbox{GeV}$ (H1) or $M_Y\!<\!2.3\ \mbox{GeV}$ (ZEUS). The cut on $t,$ the 
squared momentum transfer at the proton vertex, is $|t|\!<\!1\ \mbox{GeV}^2$ for 
both experiments. These data are presented in terms of the $t$-integrated 
reduced cross section $\sigma_r^{D(3)}(\xpom,x,Q^2)$ or the diffractive 
structure function $F_2^{D(3)}(\xpom,x,Q^2)$ obtained from the relations
\be
\frac{d^3 \sigma^{ep\rightarrow eXY}}{d\xpom\ dx\ 
dQ^2}=\f{4\pi\alpha_{em}^2}{xQ^4}
\lr{1-y+\f{y^2}{2}}\sigma_r^{D(3)}(\xpom,x,Q^2)
\hspace{1cm}
\sigma_r^{D(3)}=F_2^{D(3)}-\f{y^2}{1+(1-y)^2}F_L^{D(3)}
\ee
such that $\sigma_r^{D(3)}\!=\!F_2^{D(3)}$ is a very good approximation, except 
at large $y\!=\!W^2/s$ with $s$ the total energy in the $e\!-\!p$ collision. H1 
and ZEUS measurements are realised with different $M_Y$ cuts so the two 
experiments do not measure exaclty the same cross-section: the 
proton-dissociative events are reduced in the range of the H1 data set. However 
the difference is a known constant factor: ZEUS data points can be converted to 
the same $M_Y$ range as H1 by multiplying ZEUS values by the factor 
$0.85$~\cite{f2d97,zeus}. There exist also data from ZEUS for which the proton 
has been detected in the final state~\cite{zeuslps} that we include in the 
following analysis. These data correspond to our definition of diffractive 
events given in Section II, as the proton truelly escapes the collision intact. 
But again, the obtained cross-section differs by only a constant factor from the 
one measured without tagging the final-state proton. To be comparable with the 
H1 data~\cite{f2d97}, we multiply the data in~\cite{zeuslps} by the factor 
$1.23.$

In order to test the geometric scaling properties exhibited above, we
first express $d\sigma^{\g^*p\rightarrow Xp}_{diff}/d\beta$ in terms of
the diffractive structure function:
\be
\beta\ \f{d\sigma^{\g^*p\rightarrow Xp}_{diff}}{d\beta}
=\f{4\pi^2\alpha_{em}}{Q^2} \xp F_2^{D(3)}\ .\ee
In Fig.2, we present the measurements of the H1~\cite{f2d97} and 
ZEUS~\cite{zeus,zeuslps} collaborations for $\beta\ d\sigma^{\g^*p\rightarrow 
Xp}_{diff}/d\beta$ as a function of $\tau_d\!=\!Q^2/Q_s^2(\xp)$ at six fixed 
value of $\beta:$ 0.04, 0.1, 0.2, 0.4, 0.65 and 0.90. For each of them, we 
include all data points for $Q^2$ values in the range $[5;90]\ \mbox{GeV}^2$ and 
for $\xpom\!<\!0.01.$ For the ZEUS data points, the bin-center values in $\beta$ 
given in~\cite{zeus,zeuslps} are not exactly the ones quoted on Fig.2. To be 
able to compare H1 and ZEUS data sets, we have extrapolated ZEUS data points to 
the closest H1 bin center in $\beta$. The correction is obtained from a BEKW 
fit~\cite{zeus} on the ZEUS data sets~\cite{zeus,zeuslps}. We have used the 
saturation scale \eqref{qs}. It is clear on Fig.2 that the HERA experimental 
measurements of the diffractive cross-section in DIS are compatible with the 
geometric scaling property predicted by formula \eqref{test1}, as for each 
$\beta$ bin, the different points form a line.
\begin{figure}[ht]
\begin{center}
\epsfig{file=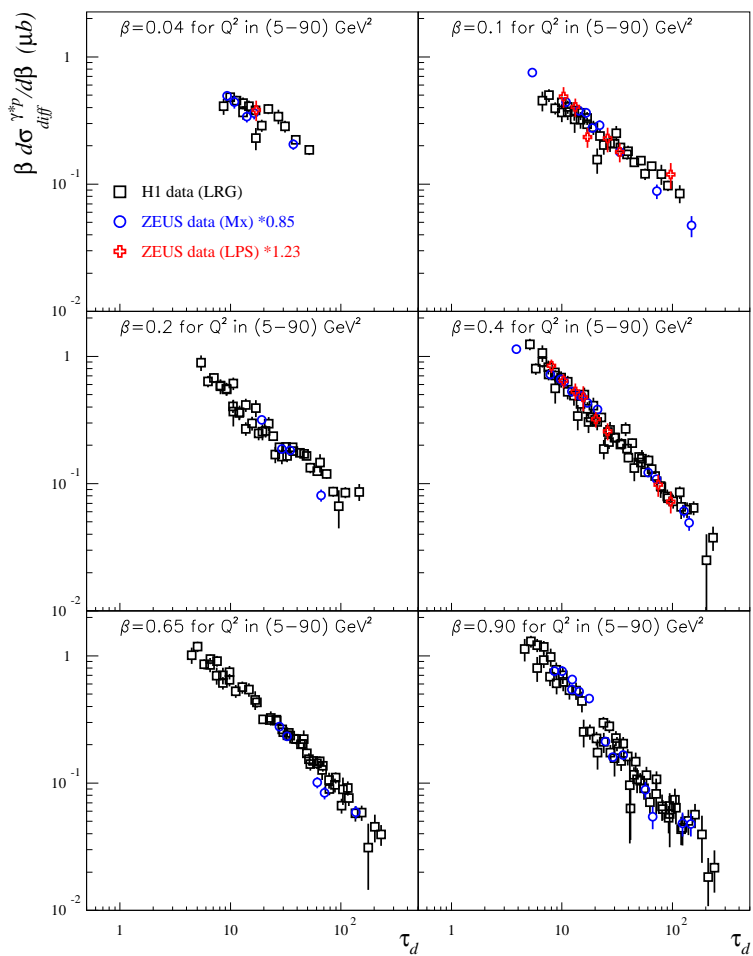,width=10cm}
\caption{The diffractive cross section
$\beta\ d\sigma^{\g^*p\rightarrow Xp}_{diff}/d\beta$ from H1 and ZEUS 
measurements, as a function of $\tau_d$ in bins of $\beta$ for $Q^2$ values in 
the range $[5;90]\ \mbox{GeV}^2$ and for $\xpom\!<\!0.01.$ Only statistical 
uncertainties are shown.}
\end{center}
\label{F2}
\end{figure}

Let us finally confront the prediction \eqref{test2} with DVCS and vector-meson 
production data from HERA. On Fig.3, the available measurements from H1 and ZEUS 
for DVCS~\cite{dvcs}, $\rho$~\cite{rho}, $\phi$~\cite{phi} and 
$J/\psi$~\cite{jpsi} exclusive productions are displayed. We represent the total 
cross-sections (meaning $t$-integrated) as a function of $\tau_V$ 
($\tau_V\!=\!\tau$ for DVCS) with the saturation scale \eqref{qs}. Again, for 
each vector meson, the data lie on a single curve, confirming the geometric 
scaling prediction.
\begin{figure}[ht]
\begin{center}
\epsfig{file=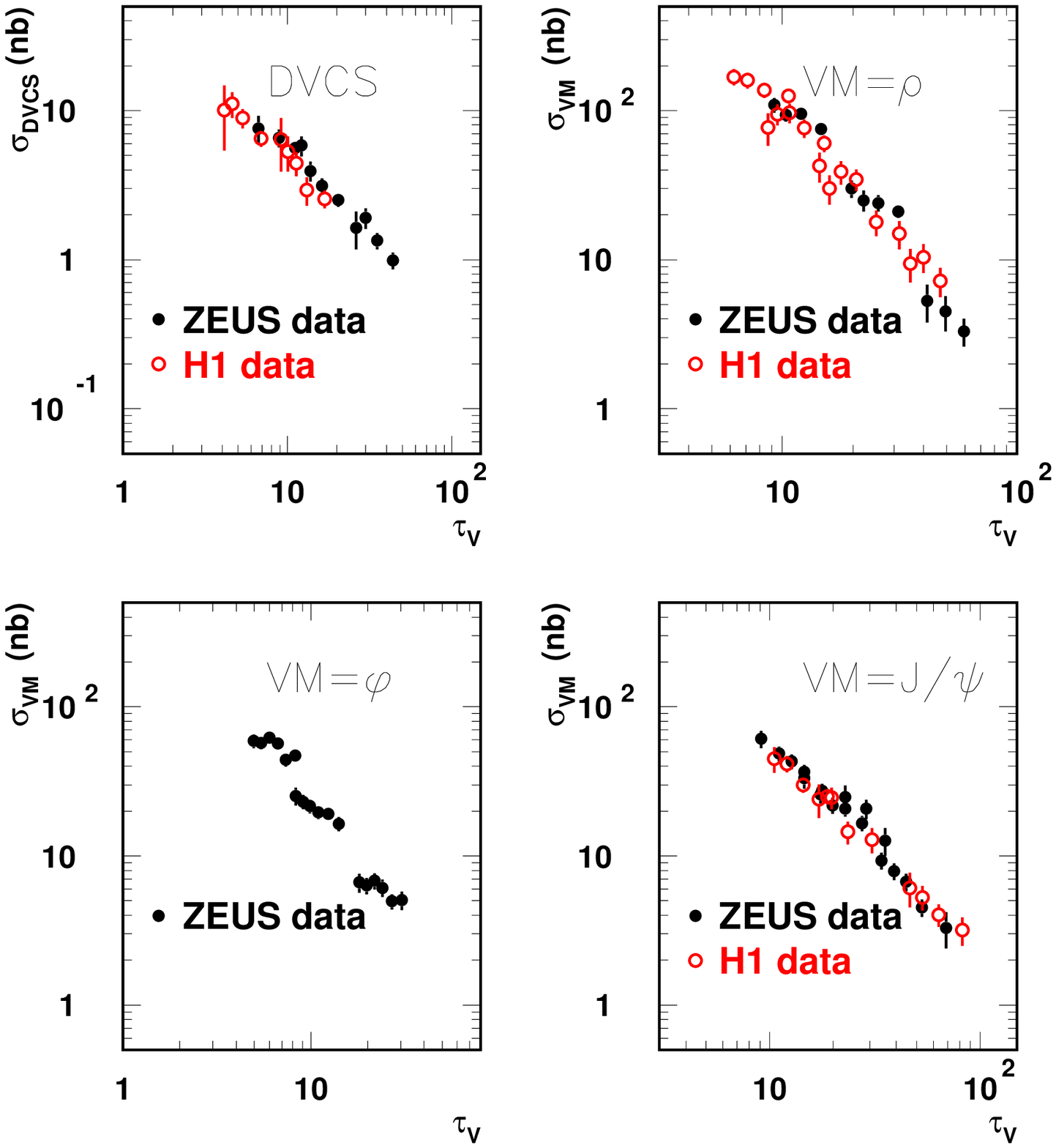,width=10cm}
\caption{The $\rho$, $J/\Psi$ and $\phi$ production cross-sections 
$\sigma^{\g^*p\rightarrow Vp}_{VM}$ and the DVCS cross-section 
$\sigma^{\g^*p\rightarrow \gamma p}_{DVCS}$ from H1 and ZEUS 
measurements, as a function of $\tau_V$ and for $\xpom\!<\!0.01.$ Only 
statistical uncertainties are shown.}
\end{center}
\label{F3}
\end{figure}

\section{Conclusions}
\label{sec:6}

The dipole picture of deep inelastic scattering is a theoretical framework 
provided by perturbative QCD ($Q^2\!\gg\!\Lambda_{QCD}^2$) in the high-energy 
limit ($x\!\ll\!1$). It allows to express total \eqref{tot}, diffractive 
\eqref{diff} and exclusive \eqref{vm} cross-sections in terms of a single 
object: a dipole scattering amplitude off the proton. One of the main features 
of the saturation regime of QCD is the scaling law \eqref{scal} of this dipole 
amplitude. The resulting consequence, the geometric scaling \eqref{test0} of the 
total cross-section in DIS, is well-known and geometric scaling has been found 
in the data five years ago~\cite{geoscal}. As shown in Fig.1, the present data 
confirm it.

In this letter, we have exhibited the consequences of the scaling \eqref{scal} 
for diffractive observables in DIS, namely for the diffractive cross-section 
\eqref{diff} and the vector-meson production (and DVCS) cross-section 
\eqref{vm}. We have shown that further manifestations of the saturation scale 
should appear in the diffractive data in the form of the scaling laws 
\eqref{test1} and \eqref{test2} with the same saturation scale $Q_s$ as in the 
inclusive case \eqref{test0}. We have analysed the present data and shown them 
in Fig.2 and Fig.3. These confirm the expected behaviors and suggest that all 
three scaling displayed in Fig.1, Fig.2 and Fig.3 are indeed manifestations of 
the saturation regime of QCD.

For the saturation scale $Q_s,$ we have taken over the same values of the 
parameters $\lambda$ and $x_0$ as in~\cite{golec}. We did not try to vary them 
to obtain a better scaling. In the context of the exact scaling law 
\eqref{scal}, a definite value for those parameters does not really make sense 
anyway. Indeed, fits of different QCD-inspired saturation 
models~\cite{golec,satmod} on $\sigma^{\g^*p\rightarrow X}_{tot}(x,Q^2)$ have 
shown that the precise values of the parameter are sensitive to whether or not 
scaling violations are included (they are also sensitive to what type of scaling 
violations are included). In any case, the parameters never differ significantly 
from those we used here.

Note finally that in the case of vector-meson production, we only looked at 
cross-sections integrated over the momentum transfer $t.$ Cross-sections 
differential with respect to $t$ can also be expressed in the dipole picture, 
with the $t$ dependence of the cross-section related to the impact parameter 
dependence of the dipole amplitude via a Fourier transform (see for 
instance~\cite{mms,kowtea}). We did not consider such observables in this work, 
but they represent natural places to look for geometric scaling properties at 
non-zero transfer as was predicted in~\cite{nfgs}. It would require to carry out 
measurements with broad ranges for the $x$ and $Q^2$ values (i.e. a large range 
for $\tau$), for different fixed values of $t.$ This may be an experimental 
challenge, but it would certainly be worthy. It would especially help our 
understanding of the impact parameter ($\textbf{b}$) dependence of the dipole 
amplitude, and our understanding of how it is mixed with the high-energy 
evolution.

\begin{acknowledgments}

We would like to thank Robi Peschanski for his useful remarks and comments on 
the manuscript.

\end{acknowledgments}
\newpage


\begin{thebibliography}{10}

\bibitem{glr} 
L.V. Gribov, E.M. Levin and M.G. Ryskin, {\it Phys. Rep.} {\bf 100} (1983) 1.

\bibitem{extscal}
E. Iancu, K. Itakura and L. McLerran, {\it Nucl. Phys.} {\bf A708} (2002) 327.

\bibitem{geoscal}
A.M. Sta\'sto, K. Golec-Biernat and J. Kwiecinski, {\it Phys. Rev. Lett.}
{\bf 86} (2001) 596.

\bibitem{golec} 
K. Golec-Biernat and M. W\"usthoff, {\it Phys. Rev.} {\bf D59} (1999) 014017;  
{\it Phys. Rev.} {\bf D60} (1999) 114023.

\bibitem{dipole}
A.H. Mueller, {\it Nucl. Phys.} {\bf B335} (1990) 115;
N.N. Nikolaev and B.G. Zakharov, {\it Zeit. f\"ur. Phys.} {\bf C49} (1991) 607.

\bibitem{nikzak}
N.N. Nikolaev and B.G. Zakharov, {\it Zeit. f\"ur. Phys.} {\bf C53} (1992) 331.

\bibitem{biapesroy}
A. Bialas and R. Peschanski, {\it Phys. Lett.} {\bf B378} (1996) 302; 
{\it Phys. Lett.} {\bf B387} (1996) 405; A. Bialas, R. Peschanski and C. Royon, 
{\it Phys. Rev.} {\bf D57} (1998) 6899; S. Munier, R. Peschanski and C. Royon,
{\it Nucl. Phys.} {\bf B534} (1998) 297.

\bibitem{diffscal}
E. Iancu, Y. Hatta, C. Marquet, G. Soyez and D.N. Triantafyllopoulos,
{\it Nucl. Phys.} {\bf A773} (2006) 95.

\bibitem{gbar}
J. Bartels, H. Jung and M. W\"usthoff, {\it Eur. Phys. J.} {\bf C11} (1999) 111.

\bibitem{gkop}
B.Z. Kopeliovich, A. Schaefer and A.V. Tarasov, {\it Phys. Rev.} {\bf D62} 
(2000) 054022.

\bibitem{gkov}
Yu.V. Kovchegov, {\it Phys. Rev.} {\bf D64} (2001) 114016.

\bibitem{gmun}
S. Munier and A. Shoshi, {\it Phys. Rev.} {\bf D69} (2004) 074022.

\bibitem{gmar}
C. Marquet, {\it Nucl. Phys.} {\bf B705} (2005) 319; {\it Nucl. Phys.}
{\bf A755} (2005) 603c; K. Golec-Biernat and C. Marquet, {\it Phys. Rev.}
{\bf D71} (2005) 114005.

\bibitem{gwus}
M. W\"usthoff, {\it Phys. Rev.} {\bf D56} (1997) 4311.

\bibitem{mwfs1}
J. Nemchik, N.N. Nikolaev and B.G. Zakharov, {\it Phys. Lett.} {\bf B341} (1994) 
228; J. Nemchik, N.N. Nikolaev, E. Predazzi and B.G. Zakharov,
{\it Zeit. f\"ur. Phys.} {\bf C75} (1997) 71.

\bibitem{mwfs2}
L. Frankfurt, W. Koepf and M. Strikman, {\it Phys. Rev.} {\bf D54} (1996) 3194. 

\bibitem{mwfs3}
H.G. Dosch, T. Gousset, G. Kulzinger and H.J. Pirner, {\it Phys. Rev.} {\bf D55} 
(1997) 2602; G. Kulzinger, H.G. Dosch and H.J. Pirner, {\it Eur. Phys. J.}
{\bf C7} (1999) 73.

\bibitem{bk}
I. Balitsky, {\it Nucl. Phys.} {\bf B463} (1996) 99;
{\it Phys. Lett.} {\bf B518} (2001) 235;
Yu.V. Kovchegov, {\it Phys. Rev.} {\bf D60} (1999) 034008;
{\it Phys. Rev.} {\bf D61} (2000) 074018.

\bibitem{jimwlk}
J. Jalilian-Marian, A. Kovner, A. Leonidov and H. Weigert, {\it Nucl. Phys.} 
{\bf B504} (1997) 415; {\it Phys. Rev.} {\bf D59} (1999) 014014;
J. Jalilian-Marian, A. Kovner and H. Weigert, {\it Phys. Rev.} {\bf D59} (1999) 
014015; E. Iancu, A. Leonidov and L. McLerran, {\it Nucl. Phys.} {\bf A692} 
(2001) 583; {\it Phys. Lett.} {\bf B510} (2001) 133; E. Ferreiro, E. Iancu, A. 
Leonidov and L. McLerran, {\it Nucl. Phys.} {\bf A703} (2002) 489; H. Weigert, 
{\it Nucl. Phys.} {\bf A703} (2002) 823.

\bibitem{tw}
S. Munier and R. Peschanski, {\it Phys. Rev. Lett.} {\bf 91} (2003) 232001;
{\it Phys. Rev.} {\bf D69} (2004) 034008; {\it Phys. Rev.} {\bf D70} (2004) 
077503.

\bibitem{ploop}
A.H. Mueller and A.I Shoshi, {\it Nucl. Phys.} {\bf B692} (2004) 175; 
E. Iancu, A.H. Mueller and S. Munier, {\it Phys. Lett.} {\bf B606} (2005) 342;
A.H. Mueller, A.I Shoshi and S.M.H. Wong, {\it Nucl. Phys.} {\bf B715} (2005) 
440; E. Iancu and D. Triantafyllopoulos, {\it Nucl. Phys.} {\bf A756} (2005) 
419; {\it Phys. Lett.} {\bf B610} (2005) 253.

\bibitem{schild}
D. Schildknecht, B. Surrow and M. Tentyukov, {\it Phys. Lett.} {\bf B499} (2001) 
116; {\it Mod. Phys. Lett.} {\bf A16} (2001) 1829.

\bibitem{h1f2}
C. Adloff {\it et al.} [H1 Collaboration], {\it Eur. Phys. J.} {\bf C21} (2001) 
33.

\bibitem{zeusf2} 
J. Breitweg {\it et al.} [ZEUS Collaboration], {\it  Phys. Lett.} {\bf B487} 
(2000) 53; S. Chekanov {\it et al.} [ZEUS Collaboration], {\it Eur. Phys. J} 
{\bf C21} (2001) 443.

\bibitem{e665f2} 
M.R. Adams {\it et al.} [E665 Collaboration], {\it Phys. Rev.} {\bf D54} (1996) 
3006.

\bibitem{nmcf2} 
M. Arneodo {\it et al.} [NMC Collaboration], {\it Nucl. Phys.} {\bf B483} (1997) 
3. 

\bibitem{gonmac}
V.P. Goncalves and M.V.T. Machado, {\it Phys. Rev. Lett.} {\bf 91} (2003) 
202002.

\bibitem{mms}
S. Munier, A.M. Stasto and A.H. Mueller, {\it Nucl. Phys.} {\bf B603} (2001) 
427.

\bibitem{f2d97}
A. Aktas {\it et al.} [H1 Collaboration], ``Measurement and QCD Analysis of the 
Diffractive Deep-Inelastic Scattering Cross Section at HERA'', 
arXiv:hep-ex/0606004.

\bibitem{zeus}
S. Chekanov {\it et al.} [ZEUS Collaboration], {\it Nucl. Phys.} {\bf B713} 
(2005) 3.

\bibitem{zeuslps}
S. Chekanov {\it et al.} [ZEUS Collaboration], {\it Eur. Phys. J.} {\bf C38}
(2004) 43.

\bibitem{dvcs}
A. Aktas {\it et al.} [H1 Collaboration], {\it Eur. Phys. J.} {\bf C44} (2005) 
1; S. Chekanov {\it et al.} [ZEUS Collaboration], {\it Phys. Lett.} {\bf B573} 
(2003) 46.

\bibitem{rho}
C. Adloff {\it et al.} [H1 Collaboration], {\it Eur. Phys. J.} {\bf C13} (2000) 
371; J. Breitweg {\it et al.} [ZEUS Collaboration], {\it Eur. Phys. J.}
{\bf C6} (1999) 603.

\bibitem{phi}
S. Chekanov {\it et al.} [ZEUS Collaboration], {\it Nucl. Phys.} {\bf B718} 
(2005) 3.

\bibitem{jpsi}
A. Aktas {\it et al.} [H1 Collaboration], ``Elastic J/Psi Production at HERA'', 
arXiv:hep-ex/0510016; S. Chekanov {\it et al.} [ZEUS Collaboration], {\it Nucl. 
Phys.} {\bf B695} (2004) 3.

\bibitem{satmod}
E. Iancu, K. Itakura and S. Munier, {\it Phys. Lett.} {\bf B590} (2004) 199;
J. Bartels, K. Golec-Biernat and H. Kowalski, {\it Phys. Rev.} {\bf D66} (2002) 
0114001.

\bibitem{kowtea}
S. Munier and S. Wallon, {\it Eur. Phys. J.} {\bf C30} (2003) 359; H. Kowalski 
and D. Teaney, {\it Phys. Rev.} {\bf D68} (2003) 114005.

\bibitem{nfgs}
C. Marquet, R. Peschanski and G. Soyez, {\it Nucl. Phys.} {\bf A756} (2005) 399;
C. Marquet and G. Soyez, {\it Nucl. Phys.} {\bf A760} (2005) 208.

\end{thebibliography}
\end{document}